\documentclass[pre,aps,showpacs,twocolumn]{revtex4}
\usepackage{amsfonts}
\usepackage{amsmath}
\usepackage{amssymb}
\usepackage{graphicx}
\usepackage{epsfig}
\usepackage{bm}
\begin{document}

\title{Integrating Random Matrix Theory Predictions with Short-Time Dynamical Effects in Chaotic Systems}

\author{A. Matthew Smith and Lev Kaplan}%
\affiliation{Department of Physics, Tulane University, New Orleans, Louisiana 70118}%

%\date{\today}

\begin{abstract}
We discuss a modification to Random Matrix Theory eigenstate statistics,
that systematically takes into account the non-universal short-time behavior of chaotic systems.
The method avoids diagonalization of the Hamiltonian, instead requiring only a knowledge of short-time
dynamics for a chaotic system or ensemble of similar systems.
Standard Random Matrix Theory and semiclassical predictions are recovered in
the limits of zero Ehrenfest time and infinite Heisenberg time, respectively.
As examples, we discuss wave function autocorrelations and cross-correlations,
and show that significant
improvement in accuracy is obtained for simple chaotic systems where comparison can be made with brute-force diagonalization.
The accuracy of the method persists even when the short-time dynamics of the system or ensemble is known only in a classical approximation. Further improvement in the rate of convergence is obtained when the method is combined with the correlation function bootstrapping approach introduced previously.
\end{abstract}

\pacs{05.45.Mt, 03.65.Sq}

\maketitle

\section{Introduction}

The statistical behavior of chaotic wave functions has been a key topic of investigation from the early history of quantum chaos and wave chaos physics, and its study is essential for improved understanding of resonances, transport, and long-time dynamics in non-integrable systems~\cite{mirlin00}. The standard method of determining the statistics of the eigenstates and eigenvalues of any physical Hamiltonian is by direct diagonalization. On the other hand, the semiclassical ergodic hypothesis~\cite{voros} suggests that the statistics of chaotic eigenstates in the semiclassical limit are universal: just as a typical trajectory in a classically chaotic system uniformly covers the entire energy hypersurface in phase space, independently of the detailed dynamics, so should a typical eigenstate in the corresponding quantum system be fully spread over the available Hilbert space. Such behavior follows naturally from Random Matrix Theory (RMT)~\cite{mehta}, which describes a statistical ensemble of Gaussian-random Hamiltonians having no preferred basis. Within the RMT approximation, eigenstates are random vectors either in the full Hilbert space or in the subspace given by energy and other conservation laws. For a quantum particle in a slowly-varying potential, a typical wave function then behaves locally like a random superposition of plane waves of fixed wave number, as discussed by Berry~\cite{berry77}. RMT wave function statistics have been well confirmed as a good zeroth-order approximation for a variety of physical systems in the absence of integrability.

The related ``quantum chaos conjecture''~\cite{cvg,berry81,bgs} which states that the {\it spectra} of individual classically chaotic systems also obey RMT statistics in the semiclassical limit, is similarly well established experimentally and numerically.  Recently, significant progress has been made in deriving this result analytically, starting form a periodic orbit representation~\cite{heusler}.

As a universal theory, RMT specifically excludes system-specific behavior associated with short-time dynamics, boundary conditions, or interactions. Well-recognized deviations from random wave function statistics are associated with boundary effects~\cite{urbina07,biesbdy}, finite system size~\cite{urbina07}, unstable periodic orbits~\cite{scar}, diffusion~\cite{mirlin00}, and two-body random interactions in many-body systems~\cite{tbre,kota}.
Similar deviations
from RMT spectral statistics have also been long recognized~\cite{svz}, and are known to arise from non-universal
short-time dynamics~\cite{berry85}.

Much progress has been made in understanding such deviations in specific situations of physical interest, for example chaotic wave function correlations in Husimi space associated with classical dynamics~\cite{schanz} and realistic mesoscopic S-matrices arising from a diffusive ray picture of wave propagation~\cite{weaver}. In particular, semiclassical methods~\cite{srednicki} have proven successful in quantifying the effects on wave functions of boundaries~\cite{urbina07,biesbdy} and periodic orbit scars~\cite{scar}. However the limit implied by semiclassical approximations may not always be achievable or relevant in describing actual experiments. For example, an analysis of electron interaction matrix elements in ballistic quantum dots shows that even for thousands of electrons in the dot, several important statistical quantities typically exceed random wave predictions by a factor of 3; for other quantities the random wave model even fails to predict the correct sign~\cite{alhassid} (see also \cite{tomsovic}). This failure of the random wave picture has important implications for interpretation of conductance experiments in the Coulomb blockade regime.

In some situations, e.g.,~\cite{alhassid}, brute force diagonalization of the Hamiltonian may be used to obtain correct statistics for the stationary or long-time behavior, but for very large Hilbert spaces, such as those that arise in many-body situations, $N \times N$ matrix diagonalization is often impractical, due to the $O(\normalfont{N}^{3})$ scaling. Even where it ``works'', diagonalization is unlikely to produce any intuition about the relevant physics, and must be repeated for each new Hamiltonian. Perturbation theory addresses this situation, but only when the change in the Hamiltonian is small compared to the mean level spacing. In fact, individual eigenstates of a chaotic Hamiltonian are highly sensitive to perturbations of the system, particularly in multi-particle systems. The {\it statistics} of such systems are much more robust and remain accurate for moderate perturbations.  For these reasons, a method of predicting wave function statistics for a family of Hamiltonians, similar in form but containing non-trivial perturbations, is highly desirable.

Here we present a system- and basis-independent means of supplementing RMT with short time dynamical information, an area in which little has been done systematically despite it being well known that RMT ignores dynamical effects.  Our method, introduced previously in abbreviated form~\cite{smithkaplan}, eliminates the need for diagonalization of the Hamiltonian, while providing greatly improved accuracy over standard RMT and semiclassical methods for finite systems with a finite Ehrenfest time. Instead of treating RMT and non-universal short-time behavior separately, we show that they can be naturally combined to produce useful quantitative predictions about wave function statistics in realistic non-integrable systems.

Below, we apply our techniques to eigenstate intensity auto-correlations and two-point correlations, as they are the simplest nontrivial moments of the eigenstates. Physically, these represent the average return probability for a given initial state and the average transport probability between a pair of states. Our method is easily extendable to other moments of the eigenstate probability distribution. Similarly, while we focus on the behavior of eigenstates in the position or Gaussian wave-packet (Husimi) representation, the method applies equally well to eigenstate statistics in the momentum basis (of interest for scattering problems) or in any other representation.

A useful way to think about the relationship between real non-integrable dynamical systems and RMT is that in a real dynamical system, a generic initial wave packet spreads (e.g., exponentially with a rate given by the maximal Lyapunov exponent $\lambda$ if the Hamiltonian is chaotic), and eventually covers the entire available Hilbert space of dimension $N$.  This occurs after the Ehrenfest time $T_E \sim \lambda^{-1} \ln N$ for a ballistic chaotic system. For a diffusive dynamics, the analogous time scale is the
Thouless time $T_{\rm th}$ (the time required to spread diffusively over the system~\cite{thoul}), which is proportional to the square of the sample size. RMT, on the other hand, assumes that this randomization takes place instantaneously, and the behavior is universal for all times $t>0$. Thus, RMT corresponds to a non-integrable system with an infinite Lyapunov exponent ($\lambda=\infty$), or zero Ehrenfest or Thouless time.

A Lyapunov exponent of infinity, as implied by RMT, represents an explosion. The initial state is instantly destroyed and randomly redistributed across the entire system after a single step regardless of the the details of the system. Clearly this is not physically realistic, but it is often a useful approximation for several reasons. Firstly, for highly chaotic systems, $\lambda$ is large, and the RMT predictions are quite accurate.  Secondly the predictions of RMT for many quantities are computationally simple. It is often preferable to accept the errors of RMT in order to avoid the computational difficulty of brute force diagonalization. Thirdly, in the case of {\it spectral} statistics, RMT fails only on large energy scales (corresponding to times shorter than $T_E$ or $T_{\rm th}$), and it is entirely justified to apply RMT inside energy windows of size less than the Thouless energy $T_c \sim \hbar/T_{\rm th}$ (e.g., in the study of level spacing statistics). The situation with wave functions is not so simple, as short-time dynamical effects leave an imprint at arbitrarily long times, and inevitably lead to large deviations from RMT not only for energy-averaged quasimodes but also for individual eigenstates~\cite{hellerscar}.

\section{Method}

To enable direct comparison with RMT, let us consider fully chaotic (ballistic or diffusive) dynamics without symmetry on an $N$-dimensional Hilbert space with eigenstates $|\xi_n\rangle$. To avoid ambiguities in the definition of $|\xi_n\rangle$,
we assume a non-degenerate spectrum.  Typical quantities of interest, then, are functions of the amplitudes $\langle a|\xi_n\rangle$ for any physically-motivated basis state $|a\rangle$, which may be a position or momentum state, a Slater determinant for many-body systems, or more generally an eigenstate of some zeroth-order Hamiltonian.  Given the normalization condition $\sum_{n=1}^N |\langle a|\xi_n\rangle|^2=1$, the simplest and first non-trivial moment of these amplitudes is given by the local inverse participation ratio (IPR), which measures the degree of localization at $|a\rangle$:
\begin{equation}
\label{paa}
P^{aa} \!=\! N\sum_{n=1}^N |\langle a|\xi_n\rangle|^4 \,,
\end{equation}
varying from $P^{aa}=1$ in the case of perfect delocalization to $P^{aa}=N$ for perfect localization. For two arbitrary states we similarly define
\begin{equation}
\label{pab}
P^{ab} \!= \!N\sum_{n=1}^N |\langle a|\xi_n\rangle|^2 |\langle b|\xi_n\rangle|^2 \,,
\end{equation}
a measure of intensity-intensity correlation.
These quantities are directly related to the time-averaged return probability and transport probability,
\begin{equation}
\label{paa2}
P^{aa} =\! N \lim_{T \to \infty} \frac{1}{2T}\int_{-T}^{T} dt |\langle a|a(t)\rangle|^2 
\end{equation}
and
\begin{equation}
\label{pab2}
P^{ab}= \! N \lim_{T \to \infty} {1 \over 2T}\!\int_{-T}^{T}\! dt |\langle a|b(t)\rangle|^2 \,,
\end{equation}
as may be easily verified by inserting two complete sets of states $\sum_n|\xi_n\rangle\langle\xi_n|$ in Eq.~(\ref{paa2}) or (\ref{pab2}).

In the more general case of a time-periodic Hamiltonian $H\left(t+{T_0} \right)=H(t)$, to be considered in the numerical examples below, eigenstates $|\xi_n\rangle $ will be the Floquet eigenstates, $|\xi_n({T_0})\rangle=e^{-iE_n {T_0}/\hbar}|\xi_n\rangle$, and the time integrals in Eqs.~(\ref{paa2}) and (\ref{pab2}) are replaced by discrete sums with time step $T_0$.

Obviously, higher-order moments of the eigenstate intensities $|\langle a|\xi_n\rangle|^2$, and in general the entire joint distribution of the intensities, may be considered (e.g.,~\cite{bootstrap}) in similar fashion. We may also relax the requirement that only pure states such as $|a\rangle\langle a|$ act as probes,
and instead measure the structure of chaotic eigenstates using any desired self-adjoint operator $\hat \alpha$~\cite{eckhardt}. Operator probes (of phase space size greater than or smaller than $\hbar$) will, for example, be particularly helpful in the study of hierarchical eigenstates in a mixed chaotic-regular phase space~\cite{hier}. Again, without loss of generality we may adopt the normalization ${\rm Tr} \,\hat \alpha=1$. Eqs.~(\ref{pab}) and (\ref{pab2}) become
\begin{eqnarray}
\label{palphabeta}
P^{\alpha\beta} &=& N\sum_{n=1}^N \langle\xi_n|\hat \alpha|\xi_n\rangle  \langle\xi_n|\hat \beta|\xi_n\rangle \nonumber \\
&=& N \lim_{T \to \infty} {1 \over 2T}   \int_{-T}^{T}dt\,
{\rm Tr} \,\hat \alpha \hat\beta(t) \,,
\end{eqnarray}
with the autocorrelation $P^{\alpha\alpha}$ as an obvious special case.

In the limit $N \to \infty$, where the Heisenberg time at which eigenstates are resolved becomes long compared to all other relevant time scales in the problem, averages of the form (\ref{paa})-(\ref{palphabeta}) may be obtained using short-time dynamics. [In the following, we refer to this limit as the ``semiclassical'' limit, whether or not the system Hamiltonian has a classical analogue.] Specifically, using the correct normalization for discrete-time dynamics we have
\begin{equation}
\label{pabsclimit}
\overline{P^{ab}} \approx \overline{\sum_{t=-\tau}^{\tau} P^{ab}(t)} \,,
\end{equation}
where have defined the time-domain quantity
\begin{equation}
P^{ab}(t)=|\langle a|b(t)\rangle|^2+\langle a|a(t)\rangle \langle b(t)|b\rangle \,,
\end{equation}
and where $|a\rangle$ and $|b\rangle$ may be any two states (identical, orthogonal, or in general overlapping).  The $|a\rangle=|b\rangle$ case is the special case of the IPR and we have 
\begin{equation}
P^{aa}(t)=2|\langle a|a(t)\rangle|^2\,;
\label{paat}
\end{equation}
thus Eq.~(\ref{pabsclimit}) becomes
\begin{equation}
\label{paasclimit}
\overline{P^{aa}} \approx 2\sum_{t=-\tau}^{\tau}\overline{ |\langle a|a(t)\rangle|^2} \,.
\end{equation}

In Eq.~(\ref{pabsclimit}) and in the following, $\overline{\cdots}$ indicates an ensemble average. If desired, the ensemble of Hamiltonians may be selected so that all realizations possess the same short-time dynamics $P^{ab}(t)$, in which case the average on the right hand side of (\ref{pabsclimit}) is superfluous. The cutoff time $\tau$ must be long compared to the ballistic or diffusive Thouless time \cite{thoul} (so as to include all the non-universal dynamics), and short compared to the Heisenberg time, which scales with the effective Hilbert space dimension $N$.

Importantly, no distinction is made in Eq.~(\ref{pabsclimit}) between non-universal short-time revivals that indicate deviations from RMT in the eigenstate statistics and the $O(1/N)$ short-time revivals that are present already in the context of RMT. Effectively, these random revivals are double-counted as both short-time and long-time returns. As a result, Eq.~(\ref{pabsclimit}) systematically overestimates corrections to RMT, and violates probability conservation $\sum_b P^{ab}=1$ given a complete basis $|b\rangle$ for any $\tau >0$, with the violations growing linearly as $\tau/N$. (In particular, the violation necessarily disappears in the limit $N \to \infty$, where a clean separation
of scales exists between short-time and long-time behavior.)

We now notice that all problematic aspects of Eq.~(\ref{pabsclimit}) for finite system size $N$ can be eliminated by introducing a $\tau$- and $\langle a|b\rangle$-dependent prefactor:
\begin{equation}
\label{pabfactor}
\overline{P^{ab}} \approx C_{N}^{\langle a|b\rangle}(\tau) \overline{\int_{-\tau}^{\tau} dt\, P^{ab}(t)} \,,
\end{equation}
where in particular $C_{N}^{\langle a|b\rangle}(\tau)=N/4\tau$ converges to the exact answer as $\tau \to \infty$
(similarly to Eq.~(\ref{pab2})). To fix $C_{N}^{\langle a|b\rangle}(\tau)$ for general $\tau$, we apply Eq.~(\ref{pabfactor}) to the case of RMT dynamics and obtain
\begin{equation}
\label{ratio}
\overline{P^{ab}} \approx \overline {P^{ab}_{\rm RMT}} {\overline{\int_{-\tau}^{\tau} dt\, P^{ab}(t)}
\over \;\;\;\overline{\int_{-\tau}^{\tau} dt\,P^{ab}_{\rm RMT}(t)} \;\;\;}\,.
\end{equation}
Eq.~(\ref{ratio}), and its generalizations, constitute the key result of this paper as they demonstrate the method of combining short time dynamics with RMT to produce accurate predictions for eigenstate statistics without diagonalization. It is natural to extend this approach to higher-order moments, for example,
\begin{equation}
\label{ratio3}
\overline{(P^{ab})^2} \approx \overline {(P^{ab}_{\rm RMT})^2} \; {\;\;\;\overline{\left(\int_{-\tau}^{\tau} dt\, P^{ab}(t)\right)^2}\;\;\;
\over \overline{\left(\int_{-\tau}^{\tau} dt\,P^{ab}_{\rm RMT}(t)\right)^2} }\,.
\end{equation}
The results generalize also to the statistics of operator expectation values, e.g., $\overline{P^{\alpha\beta}}$ and $\overline{(P^{\alpha\beta})^2}$, where $P^{\alpha\beta}$ is defined by Eq.~(\ref{palphabeta}). For example,
\begin{equation}
\label{ratioalphabeta}
\overline{P^{\alpha\alpha}} \approx \overline {P^{\alpha\alpha}_{\rm RMT}} {\overline{\int_{-\tau}^{\tau} dt\, P^{\alpha\alpha}(t)}
\over \;\;\;\overline{\int_{-\tau}^{\tau} dt\,P^{\alpha\alpha}_{\rm RMT}(t)} \;\;\;}\,.
\end{equation}

The predictive power of Eqs.~(\ref{ratio}), (\ref{ratio3}), and (\ref{ratioalphabeta})
 will be demonstrated in Sec.~\ref{numerical} below.   

Thus, stationary eigenstate properties of a quantum chaotic system or ensemble of systems may be fully described by a combination of short-time dynamics for that system or ensemble,  with exact results from RMT, without any need for matrix diagonalization. Reassuringly, Eq.~(\ref{ratio}) and its generalizations yield exact results in three limits of interest.  In the RMT limit, i.e., for a system with infinite Lyapunov exponent, we have $P^{ab}(t)=P^{ab}_{\rm RMT}(t)$ and thus Eq.~(\ref{ratio}) predicts $P^{ab}=P^{ab}_{\rm RMT}$. Secondly, in the semiclassical limit $N/\tau \to \infty$, we find  ${P^{ab}_{\rm RMT}}\approx \overline{\sum_{-\tau}^{\tau} P^{ab}_{\rm RMT}(t)}$ (in the discrete-time case), and from Eq.~(\ref{ratio}) we recover Eq.~(\ref{pabsclimit}). Finally, in the limit where an infinite amount of dynamical data is available as input, $\tau \to \infty$, the prediction
of Eq.~(\ref{ratio}) becomes exact due to the identity (\ref{pab2}).

More importantly, as we will see in Sec.~\ref{numerical}, Eq.~(\ref{ratio}) and its extensions provide reliable approximations to exact diagonalization in situations quite far from any of these three limits, i.e., for finite-sized weakly chaotic systems, and where the only input is short-time dynamics on the scale of a few Lyapunov times. 

\section{Explicit Expressions: broken Time Reversal Symmetry}

The short-time overlaps $P^{ab}(t)$ needed as input to Eq.~(\ref{ratio}) may sometimes be known analytically, as in the case of periodic orbit scars, while in more general situations the short-time dynamics for a given system of interest is obtainable numerically, to any desired time scale $\tau$. The RMT factors in (\ref{ratio}) and its generalizations will depend on the specific random matrix ensemble of interest (e.g., orthogonal, unitary, or symplectic; Gaussian or circular). For a given random matrix ensemble, these factors may be calculated numerically or treated entirely analytically. 

For example, in the absence of time reversal symmetry (i.e., for the GUE or CUE ensembles), we have standard results (for arbitrary $|a\rangle$ and $|b\rangle$),
\begin{equation}
\label{pabrmt}
\overline{P^{ab}_{\rm RMT}} = \frac{N}{N+1} \left(1 +|\langle a|b\rangle|^2 \right)\,,
\end{equation}
while for a general self-adjoint operator
$\hat \alpha$ we apply spectral decomposition and obtain
\begin{equation}
\label{palphaalpharmt}
\overline{P^{\alpha\alpha}_{\rm RMT}} = \frac{N}{N+1} \left(2 \sum_i A_i^2+\sum_{i \ne j} A_iA_j \right) \,,
\end{equation}
where $A_i$ are the eigenvalues of $\hat \alpha$ ($\sum_i A_i=1$). Similar expressions exist in the presence
of time reversal symmetry, where one must be careful to distinguish the situations of time-reversal invariant and non-invariant basis states~\cite{tri}.

Similarly, RMT dynamical overlaps may be expressed exactly for any random matrix ensemble using RMT eigenstate statistics and the
RMT spectral form factor appropriate to that ensemble, e.g., 
\begin{align}
\label{formfactor}
&\overline{P^{ab}_{\rm RMT}(t)}=\frac{2}{N} \overline{P^{ab}_{\rm RMT}}+
\sum_{n \ne n'}\left(\overline{e^{i(E_{n'}-E_n)t}}\right)_{\rm RMT}  \\
&\times  \left(\overline{|\langle a|\xi_n\rangle|^2 |\langle b|\xi_{n'}\rangle|^2}+\overline{\langle a|\xi_n\rangle
\langle \xi_n|b\rangle\langle \xi_{n'}|a\rangle\langle b|\xi_{n'}\rangle}\right)_{\rm RMT} \nonumber \,.
\end{align}
For the specific case of discrete-time Floquet dynamics, described by the CUE ensemble, which will be relevant for the numerical examples in Sec.~\ref{numerical},
we have
\begin{equation}
\label{disc}
\overline{P^{ab}_{\rm RMT}(t)}= (1+|\langle a|b\rangle|^2) \times \begin{cases} 1 &  \text{for~}t= 0\\
{1+t/N \over N+1} & \text{for~} 1\le |t|\le N\\
{2 \over N+1} & \text{for~} |t| > N \end{cases} \,,
\end{equation}
and analogous expressions for self-adjoint operators are obtained by spectral decomposition, as in
(\ref{palphaalpharmt}). Results for other RMT ensembles (Gaussian or circular) are obtained by evaluating Eq.~(\ref{formfactor}) with the spectral form factor and eigenstate statistics appropriate to that ensemble.

For the discrete-time case, Eq.~(\ref{pabfactor}) takes the form
\begin{equation}
\label{pabfactordisc}
\overline{P^{ab}} \approx C_{N}^{\langle a|b\rangle}(\tau) \overline{\sum_{-\tau}^{\tau} P^{ab}(t)} \,.
\end{equation}
Now, substitution of Eqs.~(\ref{pabrmt}) and (\ref{disc}) into Eq.~(\ref{ratio}) yields the explicit closed-form result
\begin{equation}
\label{closedform}
\overline{P^{ab}} \approx \overline{\sum_{-\tau}^{\tau} P^{ab}(t)} \times \begin{cases} 
{N \over N+1+2 \tau+(\tau^2+\tau)/N} & \text{for~} 0\le \tau\le N\\
{N \over 2 +4\tau} & \text{for~} \tau \ge N \end{cases} \,.
\end{equation}
Here we note that all dependence on the initial-state overlap $\langle a|b\rangle$ cancels in the final expression (\ref{closedform}),
so the final result has the same form for identical, overlapping, or orthogonal states $|a\rangle$, $|b\rangle$.
As $N \to \infty$ for fixed $\tau$, Eq.~(\ref{closedform}) reduces to the semiclassical
expression (\ref{pabsclimit}), while in the limit $\tau \gg N$ of complete dynamical information, 
Eq.~(\ref{closedform}) reduces to the exact answer,
\begin{equation}
\overline{P^{ab}} = \lim_{\tau \to \infty} { N\over 4 \tau}\overline{\sum_{-\tau}^{\tau} P^{ab}(t)} \,,
\end{equation}
the discrete-time form of the identity (\ref{pab2}).

\section{Model and Numerical Results}
\label{numerical}

\subsection{Model}

We now discuss a few illustrative examples, using as our model the paradigmatic example of a quantized periodically kicked Hamiltonian~\cite{kicked} on the compact toroidal phase space $(q,p) \in [-1/2,1/2) \times [-1/2,1/2)$. To enable direct comparison with RMT, we consider fully chaotic (ballistic or diffusive) dynamics without symmetry.
The time-periodic Hamiltonian is given by 
\begin{align}
\label{ham}
&H(q,p,t)=T(p)+V(q)\sum_{n=-\infty}^\infty \delta(t-nT_0) \,,
\end{align}
where without loss of generality we will set the period $T_0$ to unity. Evolution for one period defines a discrete-time map
\begin{align}
p(t+1)&=p(t)-V'(q(t)) \nonumber \\
q(t+1)&=q(t)+T'(p(t+1))\,,
\label{discmap}
\end{align}
where we have chosen to begin each period immediately before the kick. 
The kinetic term $T(p)$ and potential kick term $V(q)$ are selected to produce a perturbed Arnold's cat map~\cite{boasman},
\begin{align}
T(p)&={ m\over 2}p^2+{ K\over 4\pi^2} \cos(2 \pi p)+t(p) \nonumber \\
V(q)&=-{m \over 2}q^2-{ K\over 4\pi^2} \cos(2 \pi q)+v(q)\,,
\label{tpvq}
\end{align}
where the original (linear) Arnold's cat map is obtained for $K=t(p)=v(q)=0$.
In the following, we take $|K|<m$ to guarantee full chaos (for $t(p)=v(q)=0$).

It is easy to see that the map (\ref{discmap}) has a fixed point (i.e., a periodic orbit of period $1$) at $q=p=0$. Since the dynamics is fully chaotic,
this fixed point is necessarily unstable, and in fact a straightforward calculation shows that its instability exponent (local Lyapunov exponent) is $\lambda_0=\cosh^{-1}\left(1+(m-K)^2/2\right)$, which reduces to  $\lambda_0\approx m-K$ for $m-K \ll 1$. Thus, the perturbation parameter $K$, in addition to breaking the linearity of the original cat map, makes it easy to control the chaoticity of the dynamics.

Finally, the perturbations $t(p)$  and $v(q)$ in Eq.~(\ref{tpvq}) serve to break the parity and time reversal symmetries of the Hamiltonian. We choose these functions to be random
within a small region near the edges of the phase space ($|p|>1/2-\Delta$ and $|q|>1/2-\Delta$) and zero elsewhere. In the following, we set $\Delta=0.1$, but the results have no significant dependence on $\Delta$. Conveniently, the need for symmetry breaking simultaneously allows us easily to construct ensembles of similar Hamiltonians (having the same local Lyapunov exponent $\lambda_0$), by varying $t(p)$ and $v(q)$ in Eq.~(\ref{tpvq}) while keeping $m$ and $K$ fixed.   

Quantization of the discrete-time dynamics (\ref{discmap}) on a Hilbert space of dimension $N$
is given in the usual way~\cite{kicked,boasman} by the one-step unitary evolution operator
\begin{equation}
\hat U=e^{-i\hat T(\hat p)/\hbar}e^{-i\hat V(\hat q)/\hbar} \,,
\end{equation}
where $\hbar=1/2\pi N$, $q$ and $p$ both take on discrete values $j/N$ ($-N/2 \le j <N/2$), and the eigenstates of $\hat U$ are the Floquet eigenstates $|\xi_n \rangle$.

\subsection{Inverse Participation Ratio}

\begin{figure}[h]
{
\psfig{file=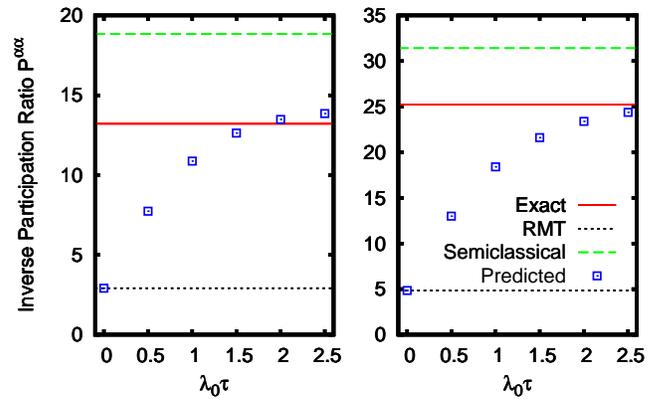,width=0.5\textwidth,angle=0}
}
	 \protect\caption{ (Color online) The inverse participation ratio $P^{\alpha\alpha}$  for a Gaussian distribution centered on
	 a short periodic orbit with local Lyapunov exponent $\lambda_0=0.5$ is computed by direct diagonalization, solid lines, and
	 compared with  the short-time dynamical prediction given by Eq.~(\ref{ratio}), squares. Here the system size is $N=32$ and the Gaussian distribution has size $s=0.5$ (Left panel) or $s=0.25$ (Right panel). Convergence of the dynamical prediction to the exact result is observed when the dynamical calculation includes information about times $\tau$ up to $2$ or $3$ in units of the local Lyapunov exponent $\lambda_0$.
The RMT value $P^{\alpha\alpha}_{\rm RMT}=(1+s^{-1})N/(N+1)$ and the semiclassical result $P^{\alpha\alpha}_{\rm SC}=(1+s^{-1})\sum_{t=-\infty}^{\infty} \text{sech}(\lambda_0 t)$ are shown for comparison.	 
	 All quantities appearing here and in subsequent figures are dimensionless.}
		\label{fig_ipralphat}
\end{figure}
	
We begin by considering the inverse participation ratio $P^{\alpha\alpha}$, where $\hat \alpha$ is the Weyl transform of a Gaussian distribution $\rho(q,p)\sim e^{-q^2/\sigma_q^2-p^2/\sigma_p^2}$ centered on the periodic orbit $q=p=0$. We define $s=\sigma_q\sigma_p/\hbar$ as the size of the Gaussian distribution.
In the special case $s=1$, $\hat \alpha$ is a projection onto a minimum uncertainty Gaussian wave packet, while more generally $\hat \alpha$ represents a mixed initial state,
corresponding to a classical phase space area smaller than or greater than $\hbar$.

Results are shown in Fig.~\ref{fig_ipralphat}, where the dynamical prediction of Eq.~(\ref{ratio}) for several values of the cutoff time $\tau$ is compared with exact values obtained by brute-force diagonalization, and also compared with the semiclassical and RMT approximations. We note that the dynamical prediction is identical to the RMT limit for $\tau=0$, as it must be since the ratio of the integrals in Eq.~(\ref{ratio}) is $1$ at $\tau=0$ and the method thus reduces to the RMT prediction when no dynamical information is available. We also notice that the prediction for the inverse participation ratio quickly converges to the exact answer after only $2$ or $3$ Lyapunov times.

\begin{figure}[h]
\centerline{\includegraphics[width=0.43\textwidth,angle=0]{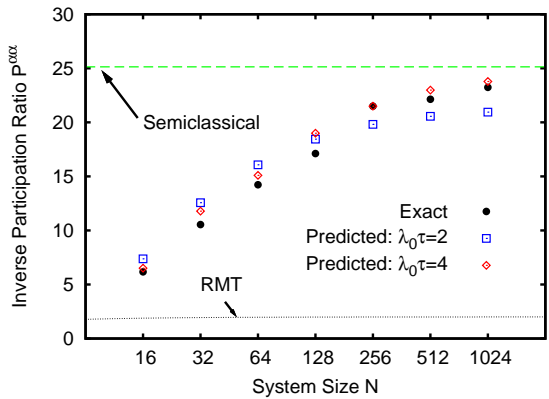}}
	 \protect\caption{ (Color online)
The inverse participation ratio for a pure Gaussian wave packet ($s=1$) is computed exactly and compared with the dynamical prediction of Eq.~(\ref{ratio}) using dynamical information up to times $\tau=2\lambda_0^{-1}$ and $4\lambda_0^{-1}$, where
$\lambda_0=0.25$ is the local Lyapunov exponent. Results are shown for various values of the system size $N$. The semiclassical and RMT limits are also shown for comparison (see Fig.~\ref{fig_ipralphat} caption).}
		\label{fig_ipralpha}
\end{figure}
Fig.~\ref{fig_ipralpha} illustrates the relationship between the exact value of the inverse participation ratio, our dynamical prediction, and the limiting RMT and semiclassical approximations, as the system size $N$ is varied. The RMT
approximation fails entirely for all system sizes except the trivial case $N=1$ (not shown).
The semiclassical limit is obtained as $N \to \infty$, but we note very significant deviations from this limit even when $N$ takes values of $100$ or greater. The strong deviations from both RMT and semiclassical predictions are well reproduced in our dynamical calculation for all system sizes $N$, even using a short calculation time $\lambda_0 \tau=2$.

\begin{figure}[h]
\centerline{\includegraphics[width=0.33\textwidth,angle=0]{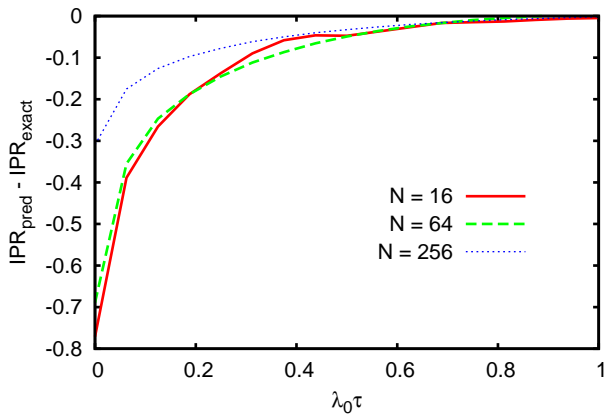}}
	 \protect\caption{ (Color online)
The inverse participation ratio in the position basis for $\lambda_0=0.0625$ is computed exactly and compared with the dynamical prediction of Eq.~(\ref{ratio}). Results are shown for several system sizes $N$ with the short time dynamics carried out to various times $\tau$.}
		\label{fig_iprvlt2}
\end{figure}

The applicability of the method is in no way limited to Gaussian test states. In Fig.~\ref{fig_iprvlt2} we compare
the predicted and actual values of the inverse participation ratio in position space, $N^{-1}\sum_q P^{qq}$, where the sum is over all $N$ position states $|q\rangle$. The calculation is performed for various system sizes $N$ and short-time cutoffs $\tau$. The local Lyapunov exponent is chosen to have a small value, $\lambda_0=0.0625$, so that we are very far from the RMT limit. Furthermore, examples of relatively small $N$ are included, where the system is very far from the semiclassical limit as well. As seen in Fig.~\ref{fig_iprvlt2}, excellent convergence to the exact results obtained by diagonalization is attained in only a fraction of a Lyapunov time, $\tau \approx 0.5 \lambda_0^{-1}$.

\subsection{Classical Dynamics Approximation}

In all the above examples, we have considered situations where the quantum dynamics up to some short time $\tau$ are assumed to be known exactly, and can be used to calculate properties of the eigenstates with arbitrarily good accuracy as $\tau \to \infty$. In practice, however, it is commonly the case that even the short-time behavior is known only in some approximation. For example, we may not have access to the dynamics under the full Hamiltonian $H$ but only to the dynamics under some simpler Hamiltonian $H_0$. Alternatively, we may be interested in the eigenstate statistics for an ensemble of Hamiltonians $H_0+\delta H$, where the short-time dynamics is dominated, but not completely determined, by $H_0$. Finally, in many physical situations of interest, the ``true'' system Hamiltonian is not known at all, e.g., in mesoscopic experiments where the details of individual eigenstate structure change completely from one identically prepared sample to the next. In all these situations, we would like to know the robustness of Eq.~(\ref{ratio}) and its generalizations, i.e., how well can we predict the statistical properties of eigenstates given only {\it approximate} knowledge of short-time dynamics? 

\begin{figure}[h]
\centerline{\includegraphics[width=0.33\textwidth,angle=270]{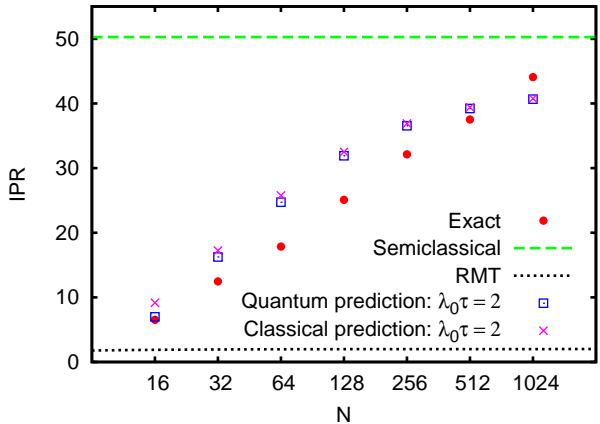}}
	 \protect\caption{ (Color online)
The inverse participation ratio for a pure Gaussian wave packet ($s=1$) is computed exactly and compared with the dynamical prediction of Eq.~(\ref{ratio}) using either exact quantum dynamical information $P^{aa}(t)$ or the classical evolution $P^{aa}_{\rm clas}(t)$ up to times $\tau=2\lambda_0^{-1}$. Here
$\lambda_0=0.125$ is the local Lyapunov exponent, and results are shown for various values of the system size $N$. The semiclassical and RMT limits are also shown for comparison (see Fig.~\ref{fig_ipralphat} caption).
}
		\label{fig_clas}
\end{figure}

In Fig.~\ref{fig_clas}, we consider a situation similar to that presented in Fig.~\ref{fig_ipralpha}, but where the short-time dynamical overlaps $P^{aa}(t)$ for $0 \le t \le \tau$ are calculated either exactly, as in Eq.~(\ref{paat}), or using classical evolution, $P^{aa}_{\rm clas}(t)=2 \iint dq \, dp \, \rho(q,p,t)  \rho(q,p)$, where $\rho(q,p)$ is the phase space density corresponding to the wave packet $|a\rangle$. The predictive power of the
classical short-time dynamics is almost as good as that of the exact quantum short-time dynamics for all system sizes $N$ considered, once correct normalization by the RMT factors is implemented, as prescribed by Eq.~(\ref{ratio}). In each case, a short-time cutoff $\tau$ equal to twice the local Lyapunov time is used for both quantum and classical dynamics; obviously further improvement in the predictive power would be obtained by using larger $\tau$. Once again, both the semiclassical and RMT approximations work very poorly for most system sizes of interest.

\subsection{Interaction Matrix Element Variance}

As another example, we consider wave function intensity correlations $P^{ab}$ for position states $|a\rangle$, $|b\rangle$. Since  ${1 \over N^2}\sum_{a,b=1}^N \overline{ P^{ab}}=1$ is given identically by wave function normalization, we focus on the first interesting moment, the variance
\begin{equation}
\label{v}
W={1 \over N^2} \sum_{a,b=1}^N \overline{ (P^{ab})^2}-1\,.
\end{equation}
$W$ is a simple measure of non-uniformity in infinite-time transport~\cite{wqe}, and ranges from $W=0$ for perfect ergodicity to $W=N-1$ for perfect localization.

We note also that we can interchange the roles of evolution eigenstates and basis states, $|\xi_n\rangle \leftrightarrow |a\rangle$.   $W$ may then be equivalently written as the variance of the interaction matrix elements   
\begin{equation}
M^{\xi_n\xi_{n'}}=N \sum_{a=1}^{N} |\langle a|\xi_n\rangle|^2 |\langle a|\xi_{n'}\rangle|^2 \,,
\end{equation}
which describe the intensity overlap of eigenstates $|\xi_n\rangle$ and $|\xi_{n'}\rangle$ in the chosen basis.
Note that the summation here is over the basis states $|a\rangle$, and not over the eigenstates $|\xi\rangle$ as in Eq.~(\ref{pab}).  The variance $W$ is then given by a summation over all pairs $|\xi_n\rangle$ and $|\xi_{n'}\rangle$,
\begin{equation}
\label{v2}
W={1 \over N^2} \sum_{n,n'=1}^N \overline{ (M^{\xi_n\xi_{n'}})^2}-1 \,.
\end{equation}
The statistics of such interaction matrix elements in chaotic systems frequently appear in applications ranging
from quantum dot conductance in the Coulomb blockade regime~\cite{alhassid} to controlling directional emission properties in microcavity lasers~\cite{turecistone}.

\begin{figure}[h!]
\centerline{\includegraphics[width=0.43\textwidth,angle=0]{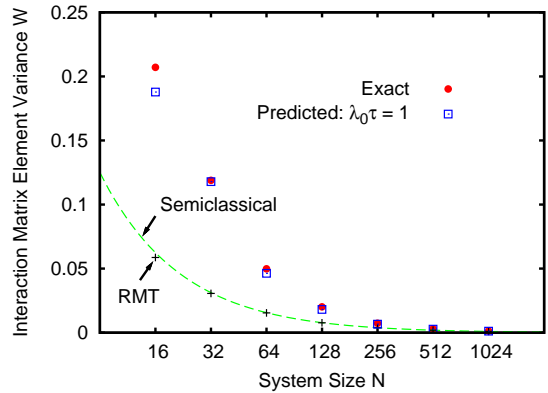}}
	 \protect\caption{ (Color online)
The interaction matrix element variance $W$ is computed exactly (Eq.~(\ref{v}) or (\ref{v2})) and compared with the short time prediction (\ref{ratio3}), with $\lambda_0\tau=1$. Here $\lambda_0=0.125$. The RMT result and the semiclassical limit $W_{\rm SC}=1/N$ are also shown for comparison.  At finite system sizes, both RMT and the semiclassical approximation fail dramatically.}
		\label{fig_ime}
\end{figure}

To predict the variance $W$ using a combination of RMT and short time dynamics, we may simply
apply Eq.~(\ref{ratio3}) to predict $\overline{ (P^{ab})^2}$ for each pair of initial and final position states $|a\rangle$ and $|b \rangle$,
and sum the results.
However, we now take the opportunity to note that the original intensity correlators $\overline{P^{ab}}$ predicted by Eq.~(\ref{ratio}) are not guaranteed to satisfy the normalization condition $\frac{1}{N^2}\sum_{a,b=1}^N \overline{P^{ab}}=1$, which holds for the exact correlators. Exact normalization is only guaranteed in Eq.~(\ref{ratio}) when $\tau \to \infty$.  This in general creates a small overall normalization error, which may affect the prediction of the higher-order moments from the short time dynamics.  However, in applying Eq.~(\ref{ratio3}) to predict $W$, we already need knowledge of the short-time dynamics, $P^{ab}(t)$, for every pair of initial and final states $|a\rangle$, $|b\rangle$. Therefore with little added computational effort we may correct for the normalization effect, further improving the convergence with $\tau$, simply by rescaling by the norm:
\begin{equation}
\label{renormal}
\overline{(P^{ab})^2} \to \frac{\overline{(P^{ab})^2}}{(\frac{1}{N^2}\sum_{a',b'=1}^N \overline{P^{a'b'}})^2} \,.
\end{equation}

Fig.~\ref{fig_ime} shows that the semiclassical and RMT predictions are very similar for the system we consider here, and both deviate significantly from the exact results for finite $N$. The method presented here, including rescaling, converges toward the exact answer very quickly, on the order of the Lyapunov time, even where the RMT and semiclassical predictions are off by a factor of 2 or 3. The accuracy can be improved further by utilizing the dynamics for longer times $\tau$.

\subsection{Bootstrapping}

Our method scales easily to arbitrarily large $N$, but for extremely large system sizes it may become time consuming to calculate even a few time steps of the evolution.  (Note that brute-force diagonalization for a system of such size is likely to be intractable.) To reduce the amount of time-domain information needed as input, one may utilize the correlation function bootstrapping approach~\cite{bootstrap}, which allows approximate evolution dynamics
to be extracted starting from as little as one unitary time step or bounce. Here we briefly review this approach,
for the simplest case of a single basis state $|a\rangle$ and discrete-time dynamics; a more general and detailed discussion is found in Ref.~\cite{bootstrap}.

Let $A^{aa}(t)=\langle a|a(t)\rangle$ be the return amplitude after time $t$. It is advantageous to partition this total return amplitude at a given time $t$ into two parts: the amplitude $A^{aa}_{\rm new}(t)$ that is returning to the subspace spanned by $|a\rangle$ for the first time at time $t$ and amplitude that has already returned to this subspace at times intermediate between $0$ and $t$. At $t=1$, we have of course $A^{aa}_{\rm new}(1)=A^{aa}(1)$. Continuing, we have
\begin{align}
\label{booting}
A^{aa}(1)=&
A^{aa}_{\rm new}(1) \nonumber \\
A^{aa}(2)=&A^{aa}_{\rm new}(1)^2+A^{aa}_{\rm new}(2) \nonumber \\
A^{aa}(3)=&A^{aa}_{\rm new}(1)^3+A^{aa}_{\rm new}(1)A^{aa}_{\rm new}(2)\\
&+A^{aa}_{\rm new}(2)A^{aa}_{\rm new}(1)+A^{aa}_{\rm new}(3) \,. \nonumber 
\end{align}
Note that in the expression for $A^{aa}(3)$, the term $A^{aa}_{\rm new}(1)A^{aa}_{\rm new}(2)$ is associated with the portion of the wave packet that leaves the initial subspace, returns after the second step and remains after the third step, while the term
$A^{aa}_{\rm new}(2)A^{aa}_{\rm new}(1)$ is associated with amplitude that remains in the initial space after the first time step, then leaves and returns again after two additional time steps. These two terms are equal in our case of a single wave packet, but would be distinct in general.

For general $t$ we can express the total return amplitude as follows,
\begin{equation}
\label{bootinggen}
A^{aa}(t)=\sum_{j=1}^t\, \sum_{t_1 +\cdots+ t_j=t} \, \prod_{k=1}^j A^{aa}_{\rm new}(t_j) \,,
\end{equation}
or equivalently
\begin{equation}
\label{booting22}
A^{aa}(t)=\sum_{t'=1}^{t} A^{aa}(t-t')A^{aa}_{\rm new}(t') \,.
\end{equation}
Inverting 
Eq.~(\ref{booting22}), we can write ${A^{aa}_{\rm new}(t)}$ as
\begin{equation}
\label{booting2}
A^{aa}_{\rm new}(t)=A^{aa}(t)-\sum_{t'=1}^{t-1} A^{aa}(t-t')A^{aa}_{\rm new}(t') \,.
\end{equation}

Now assuming the exact returns $A^{aa}(t)$ are known as before only up to time $\tau$, Eq.~(\ref{booting2}) yields the 
new retuns $A^{aa}_{\rm new}(t)$ up to time $\tau$. We now make the approximation that all new returns after time $\tau$ vanish, $A^{aa}_{\rm new}(t)=0$ for $t>\tau$, and use Eq.~(\ref{bootinggen}) or (\ref{booting22}) to calculate the total returns at arbitrary times. Obviously Eq.~(\ref{bootinggen}) or (\ref{booting22}) produces the exact total returns by construction when $t\le \tau$, while for $t >\tau$ it is only an approximation. For example, in the simplest case $\tau=1$ where only the return after one step is available as input, we have $A^{aa}(1)=A^{aa}(1)$ (exact), $A^{aa}(2 ) \approx (A^{aa}(1))^2$, $A^{aa}(3) \approx (A^{aa}(1))^3$, etc.

This bootstrapping process can be extended to an arbitrary time $\tilde \tau >\tau$, or if desired even to $\tilde \tau =\infty$. Note that the more time steps are used as input for the bootstrapping, i.e., the larger the base time $\tau$ is, the more accurate the result.

\begin{figure}[h!]
\centerline{\includegraphics[width=0.33\textwidth,angle=270]{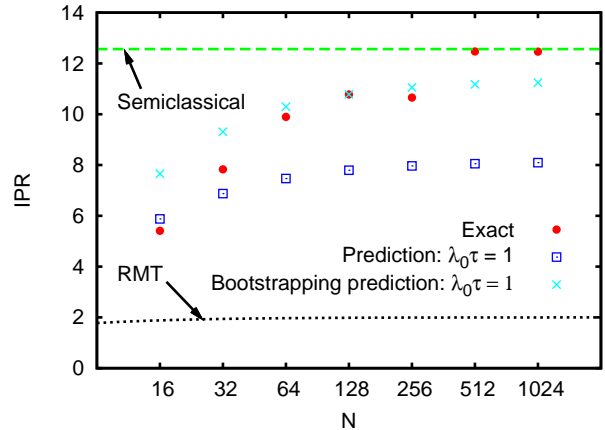}}
	 \protect\caption{ (Color online)
The ensemble averaged IPR for a minimum uncertainty wave packet, as in Fig.~\ref{fig_ipralpha}, but for an ensemble of systems with local Lyapunov exponent $\lambda_0=0.5$. Predictions with and without bootstrapping are compared to the exact IPR, to the RMT approximation, and to the semiclassical limit. Squares: exact returns up to times $\tau=2$ are used as input to predict the IPR using Eq.~(\ref{ratio}). Crosses: exact returns up to time $\tau=2$ are bootstrapped (Eq.~(\ref{bootinggen})) to extrapolate the dynamics to time $\tilde \tau=4$, and this dynamical information is then used as input to predict
the IPR using Eq.~(\ref{ratio}).  }
		\label{fig_boot2}
\end{figure}

We have found that the base time $\tau$ needed in practice is often quite small~\cite{bootstrap}, making the correlation bootstrapping approach, in combination with our dynamical method, computationally viable even for large systems. Fig.~\ref{fig_boot2} shows the results for the inverse participation ratio of a Gaussian wave packet, as in Fig.~\ref{fig_ipralpha}, but for the case $\lambda_0=0.5$, and using only two time steps of information as input, $\tau=2$. Here the squares represent
a calculation performed directly using return probabilities up to $\tau=2$ in Eq.~(\ref{ratio}), while the crosses are calculated by bootstrapping these same known recurrences up to a longer time $\tilde \tau=2 \tau$, and then inserting
return probabilities up to time $\tilde \tau$ into Eq.~(\ref{ratio}). Clearly, dynamical information
on the scale of just one Lyapunov time is in general insufficient to accurately describe eigenstate statistics, but adding bootstrapping significantly improves the accuracy, at least for moderate to large system sizes.

\section{Conclusion}

We have developed a method that improves on RMT eigenstate statistics for non-integrable systems by systematically  incorporating short-time dynamics. 
It may be applied to general quantum chaotic systems, which may or may not have a classical limit. For systems in which a classical limit exists, the approach correctly includes the effect of non-universal short periodic orbits, while simultaneously correcting for the fact that semiclassical approximations break down at low energies or small system sizes. Thus, we are able to obtain quantitative predictions both in the semiclassical
regime, and also in the regime where the semiclassical approximation fails.

The method is conceptually appealing, in that it is completely independent of the system being studied and it is also independent of the basis of interest.  In other words, while the method necessarily uses the non-universal time-domain behavior of the system as input to improve on RMT predictions, it is in no way dependent on special properties of the system for success.  Our method is computationally simpler than brute-force diagonalization, and significantly more accurate than RMT or the semiclassical limit for realistic systems.

The rate
of convergence can be improved even more using the correlation function bootstrapping  procedure. When exact short-time dynamical information is not available, approximate short-time dynamics, e.g., in a classical approximation may be successfully utilized.  The approach can be extended to consider additional eigenstate statistical measures beyond those studied here, as well as to include symmetry effects (including time reversal symmetry), mixed phase space~\cite{backer}, and resonance wave function statistics in open systems.

\begin{acknowledgments}
This work was supported in part by the NSF under Grant No.\ PHY-0545390. 
\end{acknowledgments}


\begin{thebibliography}{99}

\bibitem{mirlin00} A.\ D.~Mirlin, Phys. Rep. {\bf 326}, 259 (2000).

\bibitem{voros} A. Voros, in Stochastic Behaviour in Classical and Quantum Hamiltonian Systems, edited by
G. Casati and G. Ford, Lecture Notes in Physics Vol. 93 (Springer, Berlin, 1979), p. 326.

\bibitem{mehta} M.\ L.~Mehta, {\it Random Matrices}
(Academic Press, 2004).

\bibitem{berry77} M.\ V.~Berry, J. Phys. A {\bf 10}, 2083 (1977).

\bibitem{cvg} G.~Casati, F.~Valz-Gris, and I.~Guarneri, Lett. Nuovo Cimento {\bf 28}, 279 (1980).

\bibitem{berry81} M.\ V.~Berry, Ann. Phys. (N.Y.) {\bf 131}, 163 (1981).

\bibitem{bgs} O.~Bohigas, M.\ J.~Giannoni, and C.~Schmit, Phys. Rev. Lett. {\bf 52}, 1 (1984).

\bibitem{heusler} S.~Heusler, S.~M\"uller, A.~Altland, P.~Braun, and F.~Haake,
Phys. Rev. Lett. {\bf 98}, 044103 (2007); S.~M\"uller, S.~Heusler, A.~Altland, P.~Braun, and F.~Haake, New J. Phys. {\bf 11}, 103025 (2009).

\bibitem{urbina07} J.\ D.~Urbina and K.~Richter, Eur. Phys. J. Special Topics {\bf 145}, 255 (2007).
 
\bibitem{biesbdy} W.\ E.~Bies, N.~Lepore, and E.\ J.~Heller, J. Phys. A {\bf 36},
1605 (2003).

\bibitem{scar}	L.~Kaplan and E.\ J.~Heller, Ann. Phys. (N.Y.) {\bf 264}, 171 (1998). 

\bibitem{tbre} L.~Kaplan and T.~Papenbrock, Phys. Rev. Lett. {\bf 84}, 4553 (2000).

\bibitem{kota} V.\ K.\ B.~Kota and R.~Sahu, Phys. Lett. B {\bf 429}, 1 (1998).

\bibitem{svz} T.\ H.~Seligman, J.\ J.\ M.~Verbaarschot, and M.\ R.~Zirnbauer, Phys. Rev. Lett. {\bf 53}, 215 (1984).

\bibitem{berry85} M.\ V.~Berry, Proc. R. Soc. London A {\bf 400}, 229 (1985). 

\bibitem{schanz} H.~Schanz, Phys. Rev. Lett. {\bf 94}, 134101 (2005).

\bibitem{weaver} R.\ L.~Weaver, New J. Phys. {\bf 9}, 8 (2007).

\bibitem{srednicki} S.~Hortikar and M.~Srednicki, Phys. Rev. Lett. {\bf 80}, 1646 (1998).

\bibitem{alhassid} L.~Kaplan and Y.~Alhassid, Phys. Rev. B {\bf 78}, 085305 (2008); L. Kaplan and Y. Alhassid,
AIP Conference Proceedings {\bf 995}, 192 (2008), arXiv:0712.4095.

\bibitem{tomsovic} S.~Tomsovic, D.~Ullmo, and A.~B\"acker, Phys. Rev. Lett. {\bf 100}, 164101 (2008);
D.~Ullmo, S.~Tomsovic, and A.~B\"acker, Phys. Rev. E
{\bf 79}, 056217 (2009).

\bibitem{smithkaplan} A.\ M.~Smith and L.~Kaplan, Phys. Rev. E {\bf 80}, 035205(R) (2009).

\bibitem{thoul} D.\ J.~Thouless, Phys. Rev. Lett. {\bf 39}, 1167 (1977).

\bibitem{hellerscar} E.\ J.~Heller, Phys. Rev. Lett. {\bf 53}, 1515 (1984).

\bibitem{bootstrap} L.~Kaplan, Phys. Rev. E {\bf 71}, 056212 (2005).

\bibitem{eckhardt} B.~Eckhardt, S.~Fishman, J.~Keating, O.~Agam, J.~Main, and K.~M\"uller, Phys. Rev. E {\bf 52}, 5893 (1995).

\bibitem{hier} R.~Ketzmerick, L.~Hufnagel, F.~Steinbach, and M.~Weiss, Phys. Rev. Lett. {\bf 85}, 1214 (2000).

\bibitem{tri} L.~Kaplan, F.~Leyvraz, C.~Pineda, and  T.\ H.~Seligman,
J. Phys. A {\bf 40}, F1063 (2007). 

\bibitem{kicked}
S.~Fishman, D.\ R.~Grempel, and R.\ E.~Prange, Phys.
Rev. Lett. {\bf 49}, 509 (1982).

\bibitem{boasman} P.\ A.~Boasman and J.\ P.~Keating, Proc. R. Soc. London, Ser. A
{\bf 449}, 629 (1995).

\bibitem{wqe} L.~Kaplan and E.\ J.~Heller, Physica D {\bf 121}, 1 (1998).

\bibitem{turecistone} H.\ E.~T\"ureci, A.\ D.~Stone, and B.~Collier, Phys. Rev. A {\bf 74},
043822 (2006).  

\bibitem{backer} A.~B\"acker and R.~Schubert, J. Phys. A {\bf 35}, 527 (2002).

\end{thebibliography}
\end{document}